\documentclass[twocolumn,amsmath,amssymb]{revtex4}

\usepackage{graphicx}
\usepackage{dcolumn}
\begin{document}

\noindent{\bf A.Yurgens, D.Winkler, T.Claeson, S.Ono, and Yoichi Ando Reply to the "Comment on 'Intrinsic tunneling spectra of $\rm Bi_2(Sr_{2-x}La_x)CuO_6$' "}  

%\narrowtext

%-----------------------------------------------------------------

The Comment by Zavaritsky \cite{comment} argues that the high-bias features ("humps") that have been seen in intrinsic tunneling spectra \cite{our} can result from Joule heating and the semiconductor-like $c$-axis resistance, $R_c(T)$, of the stack. 
\begin{figure}[th]
\includegraphics[width=8cm]{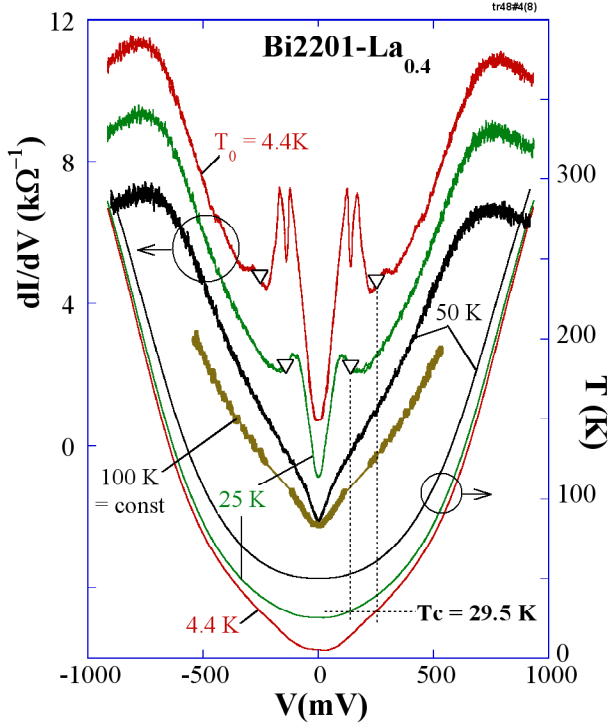}
\caption{$\mathrm{d}I/\mathrm{d}V(V)$ and $T_s(V)$ for a number of ambient temperatures $T_0$.  $\mathrm{d}I/\mathrm{d}V(V, T>10$~K) are shifted downwards (in steps of -2~k$\Omega^{-1}$) for clarity. Triangles denote data points corresponding to $T_c$. Double peaks at $T_0=4.4$~K are due to differences in the properties of the junctions in the stack. } 
\label{dIV1}
\end{figure} 

Indeed, poor thermal conductivity of all high-temperature superconductors makes them prone to local overheating. Several works used short-pulse techniques to minimize the effect of heating in stacks of intrinsic Josephson junctions (IJJ's) \cite{Gough,Suzuki}. There were attempts to directly measure the temperature of a stack utilizing its $R_c(T)\neq \mathrm{const}$ \cite{Gough2,Idriz}. In such experiments, an extra stack serving as a thermometer was placed on top of- \cite{Gough} or next to \cite{Idriz} the major stack. Due to a relatively large distance from the thermometer to the biasing contact, the temperature may not be correctly inferred. 
%under the bias lead could not be measured correctly. 
   
To practically respond to the Comment and to better address the heating issue, we measure temperature of the stacks by using micron-sized thermocouples which are in direct thermal- and electric contacts to the stacks, exactly in the place where the bias current is injected. We believe that the temperature obtained in this geometry \cite{epaps} is closest to the real one inside the stack.

Our measurements have shown that the temperature of the stack $T_s$ can reach $200-300$~K at the highest bias. For the sample shown in Fig.~1 (8$\times$8~$\mu $m$^2$ in area, with $\gtrsim$8 ITJ's), $T_s\rightarrow T_c$ at the positions of "dips" \cite{epaps}. In all cases, the pseudogap humps~\cite{our} correspond to $T_s>T_c$. Average overheating is $\sim$40~K/mW.
  
Having measured sufficiently many $I$-$V$-$T_s$ points at different bath temperatures, we can construct an isothermal, "heating-free" $\sigma(V,T_s=\mathrm{const})$ (see the curve corresponding to 100~K in Fig.~1 and Figs.~9 and 10 in \cite{epaps}). Although humps are not seen anymore at the working range of bias,  the overall non-linear shape is qualitatively preserved \cite{Suzuki,epaps}. In particular, $\sigma(V)\propto |V|$ for Bi2201 samples~\cite{our, epaps}, see Fig.~1. 

Concluding, we confirm experimentally that the Joule self-heating is a severe problem in intrinsic- and, perhaps, even in some break-junction pseudogap-spectroscopy experiments. The pseudogap features reported in our previous paper \cite{our} are indeed an artifact of Joule heating. In order to decrease the effect of heating in intrinsic junctions, one needs to drastically decrease their sizes~\cite{Krasnov_JAL}, see also Fig.~5 in \cite{epaps}.

%=======================================================
%\vspace*{3mm}
\noindent{A.Yurgens,$^1$}
{D. Winkler,$^{1,2}$}
{T. Claeson,$^1$}
{S. Ono,$^3$ and }
{Yoichi Ando$^3$}

\noindent{\small $^1$Chalmers University of Technology, 
G\"oteborg, Sweden; \\
$^2$IMEGO Institute, Aschebergsgatan 46,
G\"oteborg, Sweden; \\
$^3$Central Research Institute of Electric Power Industry (CRIEPI),
Komae, Tokyo 201-8511, Japan}   

%\vspace*{3mm}
\noindent \today

%========================================================


\begin{references}

\bibitem{comment} V. N. Zavaritsky, recent Comment, cond/mat\#0306081

\bibitem{our} A. Yurgens \textit{et al.}, Phys. Rev. Lett. \textbf{90}, 147005 (2003).

\bibitem{Gough} J. C. Fenton, \textit{et al.}, Appl. Phys. Lett. \textbf{80}, 2535 (2002).

\bibitem{Suzuki} M. Suzuki, \textit{et al.}, Phys. Rev. Lett. \textbf{82}, 5361 (1999). 

\bibitem{Gough2} C. E. Gough, \textit{et al.}, Physica C \textbf{341-348}, 1539 (2000).

\bibitem{Idriz}  I. Zogaj, Diploma work (Gothenburg, 2003), unpublished.

\bibitem{epaps} See A. Yurgens \textit{et al.}, cond/mat\#0309131 for technical details and more results and discussion. 
%\bibitem{epaps} See EPAPS Document No. [number will be inserted by publisher ] for technical details and more results and discussion. A direct link to this document may be found in the online article's HTML reference section. The document may also be reached via the EPAPS homepage (http://www.aip.org/pubservs/epaps.html) or from ftp.aip.org in the directory /epaps/. See the EPAPS homepage for more information. 

\bibitem{Krasnov_JAL} V. M. Krasnov, \textit{et al.}, J. Appl. Phys. \textbf{89}, 5578 (2001); Ibid., \textbf{93}, 1329 (2003).

\end{references}
\end{document}